\documentclass[aps,prl,reprint,superscriptaddress,floatfix]{revtex4-2}

\usepackage[ngerman,english]{babel}

\usepackage{braket}
\usepackage{amsmath}

\usepackage{graphicx}
\usepackage{floatrow}

\usepackage{changes}

\begin{document}

\title{Near-frozen non-equilibrium state at high energy in an integrable system}

\author{Stefan G.~Fischer}
\email{fischer@itp.uni-leipzig.de}
\affiliation{Institut f\"ur Theoretische Physik, Universit\"at Leipzig, Br\"uderstrasse 16, 04103 Leipzig, Germany}

\author{Yigal Meir}
\affiliation{Department of Physics, Ben-Gurion University of the Negev, Beer-Sheva, 84105 Israel}

\author{Yuval Gefen}
\affiliation{Department of Condensed Matter Physics, Weizmann Institute of Science, Rehovot, 76100 Israel}

\author{Bernd Rosenow}
\affiliation{Institut f\"ur Theoretische Physik, Universit\"at Leipzig, Br\"uderstrasse 16, 04103 Leipzig, Germany}

\date{\today}

\begin{abstract}
Ergodic many-body systems are expected to reach quasi-thermal equilibrium. Here we demonstrate that, surprisingly, high-energy electrons, which are injected into a quantum Hall edge mode with finite range interactions, stabilize at a far-from-thermalized state over a long time scale. To detect this non-equilibrium state, one positions an energy-resolved detector downstream of the point of injection.  So far, non-equilibrium distributions in integrable systems were either found not to display relaxation at all, or generically relax to near-thermal asymptotic states. In stark contrast, the here-obtained many-body state comprises fast-decaying transient components, followed by a nearly frozen distribution with a peak near the injection energy.  
\end{abstract}

\maketitle

\begin{figure}[b]
    \includegraphics[width=.95\textwidth]{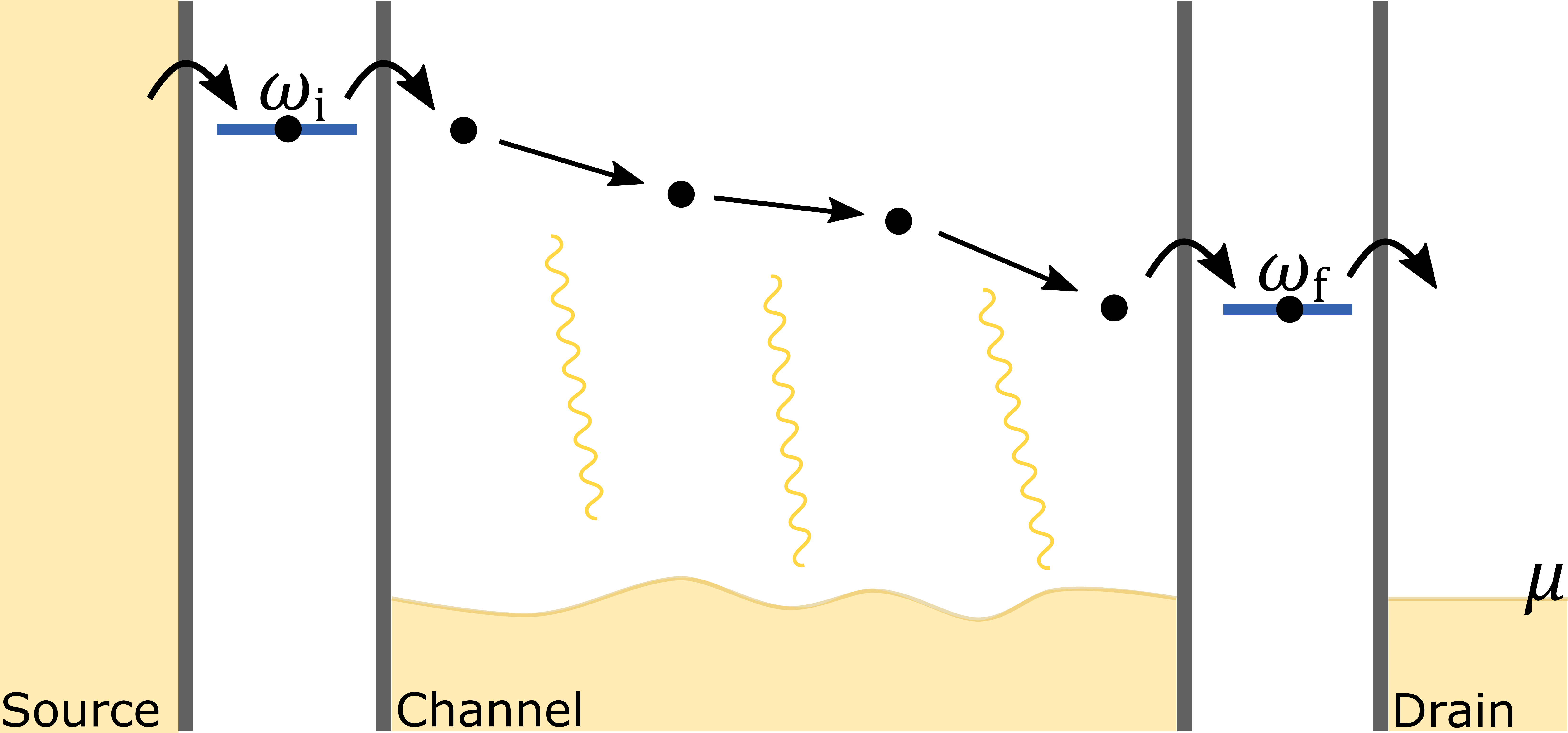}
    \caption{\label{fig:model}
A quantum dot injects electrons at energy $\omega_{\text{i}}$ into a chiral quantum channel, to subsequently be detected by a second quantum dot at energy $\omega_{\text{f}}$. 
Finite-ranged Coulomb interactions between electrons in the channel limit the average amount of  energy transferred per interaction process. 
    }
\end{figure}

Integrable many-body systems that do not thermalize challenge the universal applicability of the statistical mechanical description of large closed quantum systems in terms of thermal distributions~\cite{Polkovnikov2011,Eisert2015,Gogolin2016}.
While experimentally accessible
examples of systems that do not display ergodicity, and thereby allow to study limitations of the canonical approach,
are rare and were initially identified in
setups based on cold atomic gases~\cite{Kinoshita2006,Hofferberth2007},  
now modern condensed matter platforms provide 
precise control to explore intriguing relaxation properties, especially of the physics of integrable  one-dimensional systems.
One-dimensional quantum wires, for instance, enable the study of transport and relaxation in Luttinger liquids~\cite{Lerner2008,Calzona2017,Calzona2018,Strkalj2019}, as well as of characteristics that go beyond the Luttinger liquid paradigm~\cite{Deshpande2010,Barak2010,Idrisov2019}.
Interactions between electrons that are injected into chiral edge channels in quantum Hall systems, and electrons that copropagate in the respective channels' Fermi sea, further cause relaxation of incoming charge carrier distributions, shaped, e.g., by tunneling from quantum dots~\cite{Degiovanni2009,Ferraro2014,Acciai2017,Cabart2018,Goremykina2018,Krahenmann2019,Rodriguez2019,Rebora2021} or quantum point contacts~\cite{LeSueur2010,Degiovanni2010,Altimiras2010a,Lunde2010,Kovrizhin2011,Levkivskyi2012,Milletari2013a,Slobodeniuk2016,Schneider2017,Duprez2019,Borin2019}.

Even in the presence of strong interactions, linearity of the fermionic dispersion relation close to such channels' chemical potential leads to integrability of the dynamics. As a consequence, non-equilibrium distributions in systems that feature several channels generically relax to close-to but non-thermal metastable states~\cite{Giamarchi2004,Gutman2008,Iucci2009,Polkovnikov2011,Levkivskyi2012,Milletari2013a,Schneider2017}, which has recently been experimentally observed at intermediate propagation distances~\cite{Inoue2014,Itoh2018}. 
For larger distances, band curvature effects gain in importance and  break integrability~\cite{Imambekov2009,Barak2010,Karzig2010,Imambekov2012}. 
In such systems that feature several channels, dephasing and equilibration are adequately described by treating interactions as short-ranged, i.e.~point-like~\cite{LeSueur2010,Degiovanni2010,Lunde2010,Kovrizhin2011,Levkivskyi2012,Milletari2013a,Ferraro2014,Slobodeniuk2016,Schneider2017,Acciai2017,Cabart2018,Duprez2019,Rodriguez2019}.
For a single quantum Hall edge channel, however, point-like interactions merely cause a global renormalization of the channel's plasmon velocity, which does not cause relaxation of incoming distributions.
Finite range interactions, which give rise to a collection of distinct velocities in the channel, must be considered in order to realize equilibration~\cite{Chalker2007,Neuenhahn2008,Neuenhahn2009,Degiovanni2009,Kovrizhin2011,Cabart2018}.

\begin{figure}[t]
    \includegraphics[width=\textwidth]{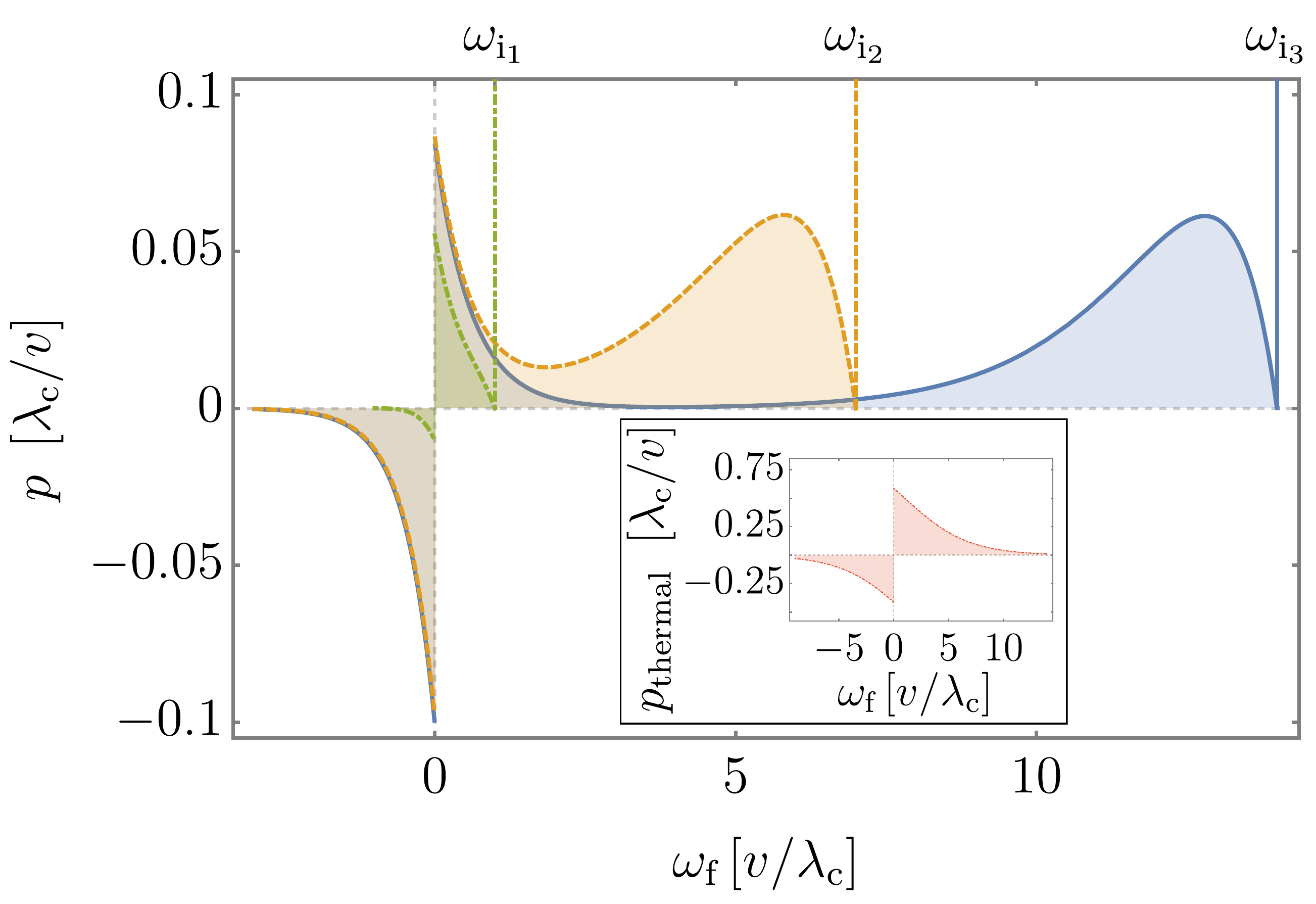}
    \caption{\label{fig:CurrentPeakPole}
    The distribution $p$ of electrons at the detection energy $\omega_{\text{f}}$ with respect to the channels ground state, after injection of an electron at energy $\omega_{\text{i}}$, is shown at fixed distance $x$ between injection and detection points. 
    Here $p$ has been obtained from bosonization [cf.~Eq.~(\ref{eq:W})] for a model that features one velocity $v$ for the injected electron and one velocity $\bar v > v$ for all plasmons in the Fermi sea [cf.~Eq.~(\ref{eq:Gvbarv})], at $\bar v = 1.2v$ and $x_{\text{s}} = 0.5\lambda_{\text{c}}$, where $\lambda_{\text{c}}$ corresponds to the screening length.
         At high injection energies ($\omega_{\text{i}_3}=14v/\lambda_{\text{c}}$, blue full), the distribution of injected electrons that dissipate some of their energy  (next to the delta function peak of elastically transferred electrons at $\omega_{\text{i}_3}$) is completely separated from the distribution of electron-hole pairs (close to $\mu = 0$). 
The distribution of originally injected electrons, see Eq.~(\ref{eq:Winjected}), here remains concentrated within $\sim\bar v / \lambda_{\text{c}}$ of the injection energy $\omega_{\text{i}_3}$, and thus highly non-thermal (compare thermal distribution obtained for $\omega_{\text{i}}=\omega_{\text{i}_3}$ and finite injection rate in inset), even in the limit of $x/\lambda_{\text{c}} \to \infty$.
    For intermediate injection energies ($\omega_{\text{i}_2}=7v/\lambda_{\text{c}}$, yellow dashed), scattering to energies in the region of energetic overlap of injected and excited electrons is suppressed by Pauli blockade. 
    At low injection energies ($\omega_{\text{i}_1}=v/\lambda_{\text{c}}$, green dot-dashed), Coulomb repulsion reduces the rate of electrons tunneling into the channel (zero-bias anomaly~\cite{Gutman2008}), and Pauli blockade further reduces relaxation.
    }
\end{figure}

\begin{figure}[t!]
    \includegraphics[width=\textwidth]{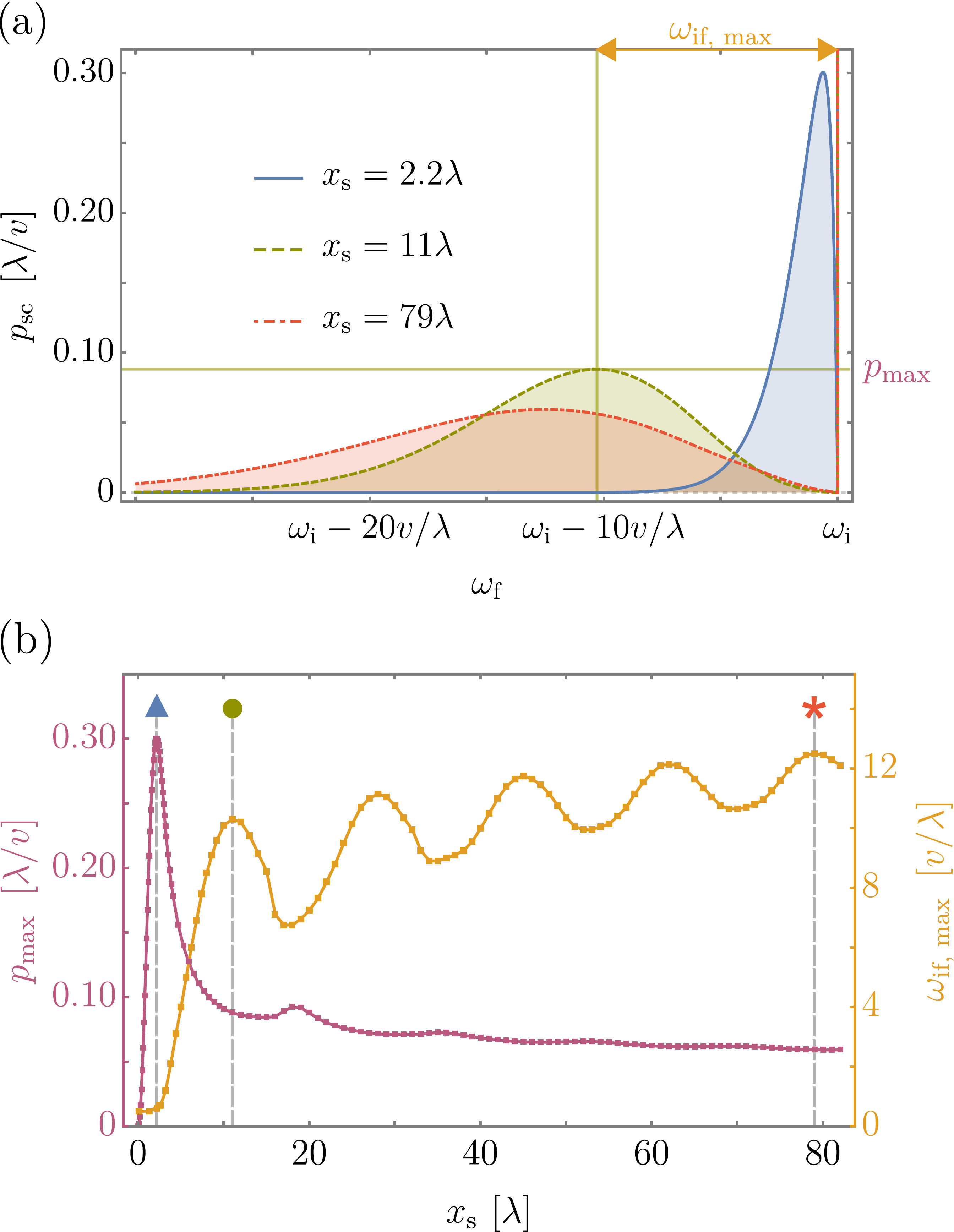}
    \caption{\label{fig:FullNumeric}
    \textbf{(a)} Distribution $p_{\text{sc}}$ of high-energy injected electrons that underwent relaxation, cf.~Eq.~(\ref{eq:pHighEexplicit}),  as a function of detection energy $\omega_{\text{f}}$, for several values of $x_{\text{s}}$. 
    Here, this distribution has been obtained for the screened interaction~(\ref{eq:nuq}), which gives rise to a continuum of plasmon velocities in the Fermi sea. 
    The energy loss $\omega_{\text{if, max}}$ (yellow arrows) at the maximum of the distribution $p_{\text{max}}$ (horizontal green line) is indicated for $x_{\text{s}} = 11\lambda$, after the distribution underwent build-up and a short period of rapid decay (green curve).
    Following this period, $p_{\text{sc}}$ varies only slowly, and remains far from thermal equilibrium.
   \textbf{(b)} Energy loss $\omega_{\text{if, max}}$ (yellow) and peak height $p_{\text{max}}$ (purple) of the distribution $p_{\text{sc}}$, as a function of $x_{\text{s}}$ [markers indicate $x_{\text{s}}$-values corresponding to distributions
    in panel (a)].
   During an initial build-up period ($x_{\text{s}} = 0$ to blue triangle), the dynamics of the electron distribution resembles the dynamics of the two-velocity model~(\ref{eq:Winjected}) [compare full blue curve at $x_{\text{s}}=2.2 \lambda$ in panel (a), and full blue curve in Fig.~\ref{fig:CurrentPeakPole}, next to $\omega_{\text{i}_3}$]. 
   In contrast to the two-velocity model, this build-up period is here followed by rapid decay (blue triangle to green circle).
   After this phase of rapid decay, the electron distribution develops a metastable form. The distribution as a function of $x_{\text{s}}$ then varies only slowly on the scale of the screening length (cf.~green circle to red asterisk), and does not display efficient decay towards the Fermi level located at $\omega_{\text{f}}=0$.  
    }
\end{figure}

In this paper, we theoretically investigate energy relaxation in such a single, one-dimensional chiral channel.
 Electrons that are injected at a well defined energy interact with the channel's Fermi sea via finite range interactions, which causes equilibration of the channel's electron distribution downstream of the injection point.
 In stark contrast to the near-thermal states mentioned above which concentrate close to the Fermi level and decay monotonously with increasing energy,
 we predict a rather extreme far-from-thermal state, in which the injected electrons remain pinned near their injection energy,  giving rise to a non-monotonous double-peak distribution.
For a simplified model that features one velocity for plasmons excited from the Fermi sea and another velocity for the injected electron, we use  bosonization to compute the full non-equilibrium electron distribution as a function of the distance between 
injection and detection of electrons. Surprisingly, the channel's full electron distribution exhibits a state far from thermal equilibrium that remains asymptotically stable [cf.~Eq.~(\ref{eq:Winjected}) and Fig.~\ref{fig:CurrentPeakPole}]. In order to test this result for a more general model of a screened interaction that 
 features a continuum of plasmon velocities, we consider the limit of high injection energy, in which the originally injected electron can be energetically 
 distinguished from Fermi sea excitations.  In this more general framework, which can be realized at the quantum Hall edge with present day experimental technology, only initially the dynamics resembles the dynamics of the above described two-velocity model. This initial period is followed by a phase of rapid decay, after which the state of the system remains metastable close to the injection energy, and thus far from thermal equilibrium (see Fig.~\ref{fig:FullNumeric}a).

In order to describe injection and detection of single electrons at specific energies in 
the chiral channel,
we consider the model depicted in Fig.~\ref{fig:model}. 
A quantum dot emits an electron at energy $\omega_\text{i}$ from a source contact into the channel at chemical potential $\mu$~\footnote{The chemical potential $\mu$ defines the reference energy which is set to zero throughout the paper.}.
This injected electron propagates along the channel for a distance $x$, before this electron itself, or electrons and holes excited during propagation, are detected by a second quantum dot at energy $\omega_\text{f}$, 
to produce a signal in the drain.   
The Hamiltonian for the chiral quantum channel is given by
\begin{align}
\label{eq:H}
H &= \int  dk \,  v k \hat{c}_k^{\dagger}  \hat{c}^{\vphantom{\dagger}}_k +  \frac{1}{4\pi} \int dk  dk^\prime  dq \, \nu_q^{\vphantom{\dagger}} \hat{c}^\dagger_{k-q} \hat{c}^\dagger_{k'+q} \hat{c}_{k'}^{\vphantom{\dagger}} \hat{c}_{k}^{\vphantom{\dagger}}  
\end{align}
in which $v$ denotes the bare velocity in the channel.
The matrix element $\nu_q$ constitutes the Fourier transform of the screened Coulomb interaction matrix element in real space, with strength $\nu_0 = \nu$ and screening length $\lambda$.
Relaxation in such channels is almost completely suppressed for injection energies $\omega_{\text{i}}$ below the quotient of the highest plasmon velocity $\bar v$ and the screening length $\lambda$~\cite{Karzig2012,Neuenhahn2008,Cabart2018, Fischer2019}. 
Below this threshold, injected electrons remain energetically indistinguishable from charge carriers excited from the Fermi sea, which causes 
Pauli blockade of relaxation.
For point-like interactions in real space ($\lambda \to 0$), the ratio $\bar v / \lambda$ diverges such that no relaxation occurs at all. Here, in the opposite limit $\omega_{\text{i}} \gg \bar v / \lambda$, we observe inhibition of relaxation that does not rely on the aforementioned effect.

In a description which features the velocity $v$ of the electron injected at high energy and one velocity $\bar v = v + \nu/2\pi$ for all plasmons excited from the Fermi sea, the full electron distribution in the channel can be obtained analytically via bosonization. This distribution is shown in Fig.~\ref{fig:CurrentPeakPole} for several values of injection energy. For high injection energies $\omega_{\text{i}} \gg \bar v / \lambda_{\text{c}}$ (blue curve), in which $\lambda_{\text{c}}$ constitutes an effective screening length of the order of $\lambda$ [cf.~Eqs.~(\ref{eq:Winjected}),~(\ref{eq:Sexplicit}), and~(\ref{eq:Gvbarv}) below],  
the injected electron can be distinguished from electrons excited from the Fermi sea (around $\mu = 0$) via the detection energy $\omega_{\text{f}}$. 
The distribution of injected electrons that dissipate some of their energy (next to the delta peak for elastically transferred electrons at $\omega_{\text{i}_3}$ in Fig.~\ref{fig:CurrentPeakPole}) 
is given by
\begin{align}
\label{eq:Winjected}
  p_{\text{inelastic}}&(\omega_{\text{f}}\gg \bar v / \lambda_{\text{c}}; \text{two-velocity model}) \nonumber \\
  =& \frac{x_{\text{s}}^2}{x_{\text{s}}^2 + \lambda_{\text{c}}^2} \frac{\lambda_{\text{c}}^2}{\bar v^2} \omega_{\text{if}} \exp \left[ -\omega_{\text{if}} \frac{\lambda_{\text{c}}}{\bar v}   \right],
\end{align}
in which $\omega_{\text{if}} = \omega_{\text{i}} - \omega_{\text{f}}$ denotes the energy loss. At high injection energies, the distance $x_{\text{s}} = \left(\bar v -v \right) x/v$ corresponds to the wave packet's spatial dispersion at the detection point. Remarkably, the state in Eq.~(\ref{eq:Winjected}) remains concentrated within an interval $\bar v/\lambda_{\text{c}}$ below the injection energy, far from thermal equilibrium, even in the limit of asymptotic propagation distance $x_{\text{s}} / \lambda_{\text{c}} \to \infty$.

From second order perturbation theory, we obtain for the inelastic distribution at high detection energies~\cite{Fischer2019}
\begin{align}
\label{eq:WinjectedPert}
        p_{\text{inelastic}}(\omega_{\text{f}}\gg \bar v / \lambda_{\text{c}}; \text{2nd order pert.})
       = \frac{x_{\text{s}}^2}{\lambda^2} \frac{\lambda^2}{v^2} \omega_{\text{if}} \frac{\nu^2_{\omega_{\text{if}}/v}}{\nu^2},
\end{align}
which breaks down at $x_{\text{s}} \approx \lambda$.
For a screened interaction that is cut off exponentially in momentum space [see Eq.~(\ref{eq:nuq}) below], the distribution in Eq.~(\ref{eq:WinjectedPert}) has the same functional energy dependence as Eq.~(\ref{eq:Winjected}) for the choice $\lambda_{\text{c}} = 2 \lambda$.
Higher order interaction terms in Eq.~(\ref{eq:Winjected}) renormalize the energy scale via $v \to \bar v$, as well as the factor that governs the distribution's $x_{\text{s}}$ dependence, which ensures applicability of Eq.~(\ref{eq:Winjected}) beyond the perturbative result. Full resummation of a
 perturbation theory 
valid at high injection energies (cf.~discussion below)
reveals that the similarity of the two-velocity model result, Eq.~(\ref{eq:Winjected}), and the perturbative result, Eq.~(\ref{eq:WinjectedPert}), stems from cancellation of all terms in which more than one plasmon is excited from the Fermi sea, with higher order in $x_\text{s}$ corrections to a term that describes excitation of one plasmon.

Taking into account screened interactions in the system, that generate a continuum of plasmon velocities, causes the above-described cancellation to no longer be perfect. 
Full resummation of the high energy electron-plasmon perturbation theory, and numerical evaluation of the thus-obtained electron distribution for the aforementioned type of interaction, constitutes our main result. 
This numerically-obtained electron distribution is displayed in Fig.~\ref{fig:FullNumeric}a, for several values of $x_\text{s}$. Initially, the dynamics of the distribution ($x_{s} = 0$ to blue triangle in Fig.~\ref{fig:FullNumeric}b) resembles the dynamics obtained from the model in which all plasmons share the same velocity $\bar v$ (blue curve in Fig.~\ref{fig:CurrentPeakPole}). This initial period is followed by a phase of rapid decay (blue triangle to green circle  in Fig.~\ref{fig:FullNumeric}b). Remarkably, after this phase of decay, the center as well as the maximum of the distribution decay only very slowly (green circle to red asterisk  in Fig.~\ref{fig:FullNumeric}b) towards the Fermi level at $\omega_{\text{f}} = 0$. Dissipation of the injected electron's energy is inhibited, such that the distribution remains metastable close to the injection energy $\omega_{\text{i}}$, and thus  far from thermal, on a length scale that far exceeds the screening length.

In the following, the theoretical background to obtain the above-described results is laid out.
Linearity of the fermionic dispersion relation in the first term of Eq.~(\ref{eq:H}) allows to obtain the greater ($+$) and lesser ($-$) Green's functions
of the channel
via bosonization~\cite{Chalker2007,Neuenhahn2008,Neuenhahn2009,Kovrizhin2011},
\begin{align}
\label{eq:GFull}
  G^{\pm}\left(x, t\right) &= G_{0}^{\pm}\left(x, t\right) e^{  S^{\pm}(x,t)}.
\end{align}
Here, the non-interacting Green's function
$G_0^{\pm}\left(x, t\right) =  1/2\pi(x - vt \pm i\epsilon)$ 
has been separated from the part that describes interactions via the exponent~\cite{Neuenhahn2008} 
\begin{align}
\label{eq:S}
  S^{\pm}(x,t) &= \int_{0}^{\infty} \frac{dq}{q} \left( e^{\mp i \left( \omega_q t - qx \right)} - e^{\mp i \left( v q t - qx \right)} \right).
\end{align}
The bosonic dispersion relation
$\omega_q = vq \left( 1 + \nu_q/2\pi v \right)$
determines the velocities of the collective plasmon modes.

Given that tunneling to and from emitter and detector quantum dots is weak, the current out of and into the drain is proportional to the distribution of electrons at propagation length $x$~\footnote{The propagation length $x$ corresponds to the separation of emitter and detector quantum dots along the edge channel that extends from negative to positive infinity.} and energy $\omega_{\text{f}}$ with respect to the ground state of the channel, provided that an electron tunnels into this channel at $\omega_\text{i}$. The general expression for this distribution is given by~\cite{Kane2003,Takei2010,Han2016}
\begin{align}
\label{eq:W}
  &p(x,\omega_{\text{i}},\omega_{\text{f}}) = \nonumber \\ 
  &\quad \frac{ v^2}{2\pi} \int_{-\infty}^{+\infty} \! dt_0  \int_{-\infty}^{+\infty} \! dt_1 \int_{-\infty}^{+\infty} \! dt_2 \,\, e^{i \omega_{\text{f}} t_0} e^{-i \omega_{\text{i}} \left( t_1 - t_2 \right)} \nonumber \\
  &\times G^{-}\left(0, t_1-t_2\right) G^{\alpha}\left(0, -t_0\right) \Big[ \Pi^{-+}(x,0,t_0,t_1,t_2) \nonumber \\ 
  &\qquad \qquad \qquad \qquad \qquad \qquad    - \Pi^{- -}(x,0,t_0,t_1,t_2) \Big],
\end{align}
where {$\alpha$ denotes the lesser component ($-$) for $\omega_f < 0$ (below the Fermi level), and the greater component ($+$) for $\omega_f > 0$ (above the Fermi level), and
$\Pi^{\beta\gamma}(x,t_0,t_1,t_2,t_3) = G^{\beta}(x,t_0-t_3)G^{\gamma}(x,t_1 - t_2)/G^{\beta}(x,t_1-t_3)G^{\gamma}(x,t_0- t_2)$.

For interactions that decay exponentially in momentum space,
\begin{align}
\label{eq:nuq}
  \nu^{\text{(exp)}}_{q} = \nu\exp\left(- \lambda |q| \right),
\end{align}
we perform an order by order integration of Eq.~(\ref{eq:S}) in an expansion in powers of $\nu / 2 \pi v$, which 
leads to
\begin{align}
\label{eq:Sexplicit}
 S^{\text{(exp)}\pm}(x,t) = \sum_{k=1}^{\infty} \frac{1}{k} \left[ \frac{\frac{\nu}{2\pi} t}{x -vt\pm i\lambda k} \right]^k.
\end{align}
Setting $\lambda k$ to $\lambda_{\text{c}}$  in the denominator of Eq.~(\ref{eq:Sexplicit}) for each $k$ generates the Green's 
functions (cf.~supplemental material~\cite{suppmat2020})  
\begin{align}
\label{eq:Gvbarv}
  G^{\pm}&\left(x, t; \text{two-velocity model}\right) \nonumber \\
 &\qquad= \frac{1}{2 \pi} \frac{1}{x - vt \pm i\epsilon} \frac{x - vt \pm i\lambda_{\text{c}}}{x - \bar vt \pm i\lambda_{\text{c}}},
\end{align}
cf.~\cite{Neuenhahn2008,Cabart2018},
which display poles determined by the velocity $v$ of high energy electrons (which corresponds to the slope of $\omega_q$ for large $q \gg 1/\lambda$) and the velocity of low energy plasmons $\bar v$ (given by the slope of $\omega_q$ at $q=0$).

Employing the thus-obtained Green's functions in Eq.~(\ref{eq:Gvbarv}), we can evaluate the relaxation distribution in Eq.~(\ref{eq:W}) analytically.  
Above the Fermi sea, $\omega_\text{f}>0$, the electron distribution of Eq.~(\ref{eq:W}) features two contributions,
 $p(\omega_{\text{f}}>0) = P_{\text{elastic}}\delta (\omega_{\text{i}}-\omega_{\text{f}}) +  p_{\text{inelastic}}\theta(\omega_{\text{i}}-\omega_{\text{f}})$. 
The first contribution, $P_{\text{elastic}}$,
has been investigated in detail~\cite{Neuenhahn2008,Neuenhahn2009}, and describes the weight of the delta peaks located at the injection energies $\omega_{\text{i}_{i}}$  
in Fig.~\ref{fig:CurrentPeakPole}.

The second 
contribution to the electron current,
$p_{\text{inelastic}}$ (for explicit expressions see supplemental material~\cite{suppmat2020}), is composed of three parts: the distribution in Eq.~(\ref{eq:Winjected}), that corresponds to the originally injected electron entering the detector at an energy other than the injection energy $\omega_{\text{i}}$, the distribution $p^{\text{e}}_{\text{exc}}$ of electrons excited from the Fermi sea entering the detector, as well as interference terms 
(cf.~the discussion based on perturbation theory~\cite{Fischer2019} which, in contrast to present results, diverges for $x_{\text{s}} \gg \lambda$). Below the Fermi sea, $\omega_f<0$, we find a contribution $p^{\text{h}}_{\text{exc}}$ of holes left behind by excited electrons $p^{\text{e}}_{\text{exc}}$, as well as interference terms.  The full electron distribution  
obtained from Eq.~(\ref{eq:W})  
is displayed in Fig.~\ref{fig:CurrentPeakPole}, for several values of the injection energy $\omega_{\text{i}}$.
In the limit of high injection energies, $\omega_{\text{i}} \gg \bar v /\lambda_{\text{c}}$ (blue curve in Fig.~\ref{fig:CurrentPeakPole}), injected and excited electrons are energetically well-separated, and all 
 above-mentioned interference terms vanish. 

To evaluate the distribution of injected electrons for arbitrary screened interactions $\nu_q$, which allows for a continuum of plasmon velocities, we separate the contribution of injected electrons that are detected close to the injection energy from charge carriers excited from the Fermi sea. In this limit, we provide a full resummation of a perturbation expansion valid at high injection energies, that treats the injected electron independently from bosonized plasmons in the Fermi sea, and fixes the injected electrons transition time at $t = x /v$. 
The approach generates  
the expression (cf.~supplemental material~\cite{suppmat2020})
\begin{align}
\label{eq:pHighEexplicit}
	&p_{\text{sc}}\left(x, \omega_{\text{if}} \right) = \nonumber \\
	& \int_{-\infty}^{\infty} \frac{dt}{2\pi} e^{i \omega_{\text{if}} t } \exp \left\{  \int_0^{\infty}  \frac{dq}{q} 4 \sin^2\left[ \frac{qx}{2} \frac{\nu_q}{2\pi v} \right] \left(e^{-i \omega_{q} t } - 1 \right) \right\},
\end{align}
in which the second term in rounded brackets corresponds to  the contribution of elastic transport in which no plasmons are excited, and the first term in Eq.~(\ref{eq:pHighEexplicit}) in rounded brackets contains the information about relaxation via excitation of any integer number $n>0$ of plasmons.

To test the validity of the high injection energy expression in Eq.~(\ref{eq:pHighEexplicit}), we  insert Eq.~(\ref{eq:Gvbarv}), which features one velocity $\bar v$ for all excited plasmons in the Fermi sea, into Eq.~(\ref{eq:pHighEexplicit}). This directly produces Eq.~(\ref{eq:Winjected}), which had initially been obtained by evaluation of the full distribution given by Eq.~(\ref{eq:W}) followed by the limit $\omega_{\text{i}}\gg \frac{\bar v}{\lambda},\, \omega_{\text{if}} $.
For exponentially decaying interactions, Eq.~(\ref{eq:nuq}), 
we evaluate Eq.~(\ref{eq:pHighEexplicit}) in a scaling limit in which $x/\lambda \to \infty$ and $\nu/2\pi v \to 0$, while the product $x_{\text{s}}/\lambda$ of these two quantities is kept constant~\footnote{For, e.g., the distribution in Eq.~(\ref{eq:Winjected}) in dimensionless form, $\bar v p_{\text{inelastic}}/\lambda$, this approximation corresponds to $\bar v \to v$, which only lowers the energy scale $\bar v /  \lambda$ of the function's dependence on $\omega_{\text{if}}$. In Eq.~(\ref{eq:pHighEexplicit}), the approximation corresponds to $\omega_q \to vq$ in the second exponential term.  } (cf.~supplemental material).

Results of this numerical evaluation are displayed in Fig.~\ref{fig:FullNumeric}. Figure~\ref{fig:FullNumeric}a shows the distribution $p_{\text{sc}}$ for several values of $x_\text{s}/\lambda$. While the blue and green curves at $x_{\text{s}} = 2.2\lambda$ and $x_{\text{s}} = 11\lambda$, respectively, are separated by a relatively short period of rapid decay, the change between this green and the red curve at $x_{\text{s}} = 79\lambda$ 
remains comparatively small 
after a time interval that exceeds the former decay period by about an order of magnitude. The slow decay of the distribution is made further apparent in Fig.~\ref{fig:FullNumeric}b, which shows the maximum $p_{\text{max}}$ of the distribution as well as the energy loss $\omega_{\text{if, max}}$ at this maximum, as a function of $x_{\text{s}}/\lambda$. The quantities $\omega_{\text{if, max}}$ and $p_{\text{max}}$, as well as the entire distribution display oscillatory behavior, which they share with the magnitude of elastic transfer $p_{\text{sc}}^{(0)}$~\cite{Neuenhahn2008,Neuenhahn2009,suppmat2020}, at a frequency that corresponds to the maximum of the Galileo transformed bosonic spectrum $\omega_q^\text{G} = \omega_q - vq$~\footnote{For $\nu^{\text{(exp)}}_{q}$~(\ref{eq:nuq}), $\omega_{q, \text{max}}^\text{G} = \nu/2\pi \lambda e$.}. 
To obtain a quantitative measure of the slow decay of $p_{\text{sc}}$ as a function of $x_{\text{s}}/\lambda$, power law fits to the values of $\omega_{\text{if, max}}$ and $p_{\text{max}}$, at those $x_{\text{s}}/\lambda$ at which $p_{\text{sc}}^{(0)}$ displays the first five local minima, show dependencies $\omega_{\text{if, max}} \sim (x_{\text{s}}/\lambda)^{0.101 \pm 0.004}$ and $p_{\text{max}} \sim (x_{\text{s}}/\lambda)^{-0.202\pm0.006}$. Fit functions different from power laws might predict asymptotic saturation of $\omega_{\text{if, max}}$ and/or $p_{\text{max}}$ at finite values.

We conclude by noting that GaAs and 
graphene are candidate materials for the observation of the above-described metastable electronic distribution. 
At $10$T, the energy gap between the $0$th and $1$st Landau levels in GaAs is about $17$meV~\cite{Beenakker1991}, and $115$meV in graphene~\cite{Song2010}. The unit of energy $\hbar v / \lambda$ for our results in Fig.~\ref{fig:FullNumeric}, at an estimated screening length of $0.5\mu$m, here amounts to $0.1$meV in GaAs and $1$meV in graphene. For both materials, the gap thus accommodates the entire energy range displayed in Fig.~\ref{fig:FullNumeric} by  a wide margin.
Quantum dot spectroscopy experiments have already been carried out in GaAs at high injection energies~\cite{Krahenmann2019,Rodriguez2019}, and  
isolation of a single channel from a multichannel system has been experimentally realized~\cite{Altimiras2010,Duprez2019,Rodriguez2019}.
In graphene, precise control of transport has recently been demonstrated in Fabry-Perot interferometers~\cite{Deprez2021,Ronen2021}.

In summary, we have investigated the relaxation of electrons injected into a one-dimensional chiral channel, which interact with charge carriers in the channel via finite-ranged interactions. For a simplified model that features one velocity of incoming electrons and one velocity for plasmons in the Fermi sea, we found a stable highly non-thermal state, concentrated close to the energy of injected electrons. For a more realistic model that features a continuum of plasmon velocities, we predict a state that remains nearly frozen close to high injection energies, far from thermal equilibrium. 

\begin{acknowledgments}
We would like to acknowledge useful discussions with Jinhong Park, Kyrylo Snizhko, and Felix Puster. Y.G.~and B.R.~acknowledge support by DFG RO 2247/11-1. Y.M.~acknowledges support
from ISF Grant No.~359/20. Y.G.~further acknowledges support by the Helmholtz
International Fellow Award, CRC 183 (project C01), and the
German-Israeli Foundation Grant No.~I-1505-303.10/2019. Y.G.~and S.G.F.~acknowledge support by the Minerva foundation.
\end{acknowledgments}

\end{document}

%\title{Supplementary Information for ``Near-frozen non-equilibrium state at high energy in an integrable system''}

% BEGIN SUPPLEMENTARY MATERIAL 
\clearpage
\pagebreak
\setcounter{page}{1}
\onecolumngrid
\widetext
\begin{center}
    \textbf{{\large Supplemental Material for: ``  Near-frozen non-equilibrium state at high energy in an integrable system   ''}}\\[1em]
    Stefan G.~Fischer,  Yigal Meir, Yuval Gefen,   and Bernd Rosenow
\end{center}
\vspace*{0.2cm}
In this supplement we include additional analysis and results characterizing  the energy relaxation of high energy electrons injected into a single chiral channel, interacting with the Fermi sea via a finite range interaction.

\section{Full expressions for electron distribution in the two-velocity model}

Below we state explicit expressions for the inelastic contributions to the electron distribution in Eq.~(6) in the main text, obtained with the Green's functions given by Eq.~(9), which feature the velocity of the injected electron $v$ and one velocity $\bar v$ for all plasmons excited from the Fermi sea. The distribution above the Fermi sea, $\omega_{\text{f}} > 0$, is composed of two terms  
$p = P_{\text{elastic}}\delta (\omega_{\text{i}}-\omega_{\text{f}}) +  p_{\text{inelastic}}\theta(\omega_{\text{i}}-\omega_{\text{f}})$.
Elastic transfer $P_{\text{elastic}}$ has been discussed in detail in~\cite{Neuenhahn2008supp,Neuenhahn2009supp}.

 The expression for inelastically transferred electrons, which are detected at an energy $\omega_{\text{f}}$  lower than the injection energy $\omega_{\text{i}}$, can be decomposed into three parts, $p_{\text{inelastic}} = p_{\text{inj}} + p^{\text{e}}_{\text{exc}} + p^{\text{e}}_{\text{int}}$. In this decomposition, $p_{\text{inj}}$, which in the limit of high injection energies corresponds to injected electrons that lose part of their energy, is given by Eq.~(2) in the main text. The contribution which in the high injection energy limit describes electrons that are excited by the energy of the injected electron, is given by
\begin{align}
\label{eq:p_excited_e}
p^{\text{e}}_{\text{exc}} (x_{\text{s}},\omega_{\text{i}},\omega_{\text{f}}) = \frac{\lambda_{\text{c}}}{\bar{v}} \frac{x_s^2 + \lambda_{\text{c}}^2 \left( 1 - \frac{\bar v}{v} \right)^2}{x_s^2 + \lambda_{\text{c}}^2 \left( 1 + \frac{\bar v}{v} \right)^2}   \left(\frac{4v}{3\bar v} + \frac{2}{3}\right) \exp \left[ - \left( \omega_{\text{f}} + \omega_{\text{i}} \right) \frac{\lambda_{\text{c}}}{\bar v} \right] \left\{ \exp \left[ - \left( \omega_{\text{f}} - \omega_{\text{i}} \right) \frac{\lambda_{\text{c}}}{\bar v} \right] -1  \right\},
\end{align}
where $x_s = {x \nu\over 2 \pi v}$. 
The remaining contribution,  
\begin{align}
\label{eq:p_interference_e}
 &\!\!p^{\text{e}}_{\text{int}}(x_{\text{s}},\omega_{\text{i}},\omega_{\text{f}}) = \nonumber \\ 
   &\frac{\lambda_{\text{c}}}{6\bar{v}}  \left( 4\frac{v}{\bar v} - 3\frac{v^2}{\bar v^2} - 1 \right) \frac{x_s^2 + \lambda_{\text{c}}^2 \left( 1 - \frac{\bar v}{v} \right)^2}{x_s^2 + \lambda_{\text{c}}^2 } \frac{x_s^2 + (2\lambda_{\text{c}})^2}{x_s^2 + \lambda_{\text{c}}^2\left( 2 -  \frac{\bar v}{v} \right)^2}  \exp \left[ - \left( \omega_{\text{f}} + \omega_{\text{i}} \right) \frac{\lambda_{\text{c}}}{\bar v} \right] \left\{ \exp \left[ - 2\left( \omega_{\text{i}} - \omega_{\text{f}} \right) \frac{\lambda_{\text{c}}}{\bar v} \right] -1  \right\} \nonumber \\
                                                                                                   		-& \frac{2\lambda_{\text{c}}}{\bar{v}}\text{Re}\Bigg(                                                          \frac{v}{\bar v}  \frac{ i \lambda_{\text{c}} \left[ x_s +i \lambda_{\text{c}} \left( \frac{\bar v}{v} - 1 \right) \right]}{x_s^2 + \lambda_{\text{c}}^2} \frac{x_s}{x_s - i \lambda_{\text{c}}} \nonumber \exp \left[-i \omega_{\text{f}} \frac{x_s}{\bar v} - \omega_{\text{i}} \frac{\lambda_{\text{c}}}{\bar v} \right] \left\{  \exp \left[+ i \left( \omega_{\text{f}} - \omega_{\text{i}} \right) \frac{x_s - i \lambda_{\text{c}}}{\bar v} \right] -1 \right\} \nonumber \\
                                                            +&\frac{v}{\bar v}   \frac{\left[ x_s + i \lambda_{\text{c}} \left( \frac{\bar v}{v} - 1 \right) \right] \left[ 2 x_s + i \lambda_{\text{c}} \frac{\bar v}{v} \right] \left[ x_s^2 + \lambda_{\text{c}}^2 \left(1 - \frac{\bar v}{v} \right)^2 \right]}{2 x_s \left[ x_s + i \lambda_{\text{c}} \frac{\bar v}{v} \right] \left[ x_s  - i \lambda_{\text{c}} \left( 2 - \frac{\bar v}{v} \right) \right] \left[x_s + i \lambda_{\text{c}} \left( 1 + \frac{\bar v}{v} \right)\right]}   \exp \left[ - \left( \omega_{\text{f}} + \omega_{\text{i}} \right) \frac{\lambda_{\text{c}}}{\bar v} \right] \!\! \! \left\{ \exp \left[ - i \left( \omega_{\text{f}} - \omega_{\text{i}} \right) \left( \frac{x_s}{\bar v} + i \frac{\lambda_{\text{c}}}{v} \right) \right] - 1 \right\} \Bigg) ,
\end{align}
 vanishes in the limit of high injection energy, $\omega_{\text{i}}\gg \bar v / \lambda_{\text{c}}$, in which injected and excited electrons can be distinguished by the detection energy.
 
Below the Fermi sea, $\omega_{\text{f}} < 0$, we decompose the distribution into two contributions $p = p^{\text{h}}_{\text{exc}}\theta(\omega_{\text{i}} + \omega_{\text{f}}) + p^{\text{h}}_{\text{int}} \theta(\omega_{\text{i}} + \omega_{\text{f}})$. The first term,
\begin{align}
\label{eq:p_excited_h}
p^{\text{h}}_{\text{exc}}(x(x_{\text{s}}),\omega_{\text{i}},\omega_{\text{f}}) = - \frac{\lambda_{\text{c}}}{\bar v} &\frac{x_s^2 + \lambda_{\text{c}}^2\left(1-\frac{\bar v}{v}\right)^2}{ x_s^2 + \lambda_{\text{c}}^2\left( 1 +\frac{\bar v}{v} \right)^2}\frac{6}{\left( 2  +  \frac{v}{\bar  v}  \right)} \exp \left[ \left( \omega_{\text{f}}-\omega_{\text{i}} \right) \frac{\lambda_{\text{c}}}{v} \right] \left\{ \exp\left[ \left( \omega_{\text{f}} + \omega_{\text{i}} \right) \frac{\lambda_{\text{c}}}{v} \right] -1 \right\}, 
\end{align}
here describes holes left by electrons excited from the Fermi sea. The second term is given by
\begin{align}
\label{eq:eq:p_interference_h}
p^{\text{h}}_{\text{int}}(x(x_{\text{s}}),&\omega_{\text{i}},\omega_{\text{f}}) = 
  - \frac{\lambda_{\text{c}}}{\bar v} \frac{x_s^2 + \lambda_{\text{c}}^2\left(1-\frac{\bar v}{v}\right)^2}{x_s^2 + \lambda_{\text{c}}^2 \frac{\bar v^2}{v^2}} \exp \left[ \left( \omega_{\text{f}}-\omega_{\text{i}} \right) \frac{\lambda_{\text{c}}}{v} \right]  \nonumber \\
\times  \Bigg[ &  \frac{( \frac{\bar v}{v}-1)}{ (1 + 2 \frac{\bar v}{v})(2 \frac{\bar v}{v})(1 +\frac{\bar v}{v})} \left\{ \exp \left[ -\left(\omega_{\text{f}} + \omega_{\text{i}} \right)\left(  \frac{\lambda_{\text{c}}}{v} +  \frac{\lambda_{\text{c}}}{\bar v} \right) \right] - 1 \right\} \nonumber  \\
 +& \frac{(3 \frac{\bar v}{v}-1)(1 - 2  \frac{\bar v}{v})}{(-2 \frac{\bar v}{v})  (1 - \frac{\bar v}{v})}\left\{ \exp\left[ \left( \omega_{\text{f}} + \omega_{\text{i}} \right)\left(\frac{\lambda_{\text{c}}}{v} - \frac{\lambda_{\text{c}}}{\bar v} \right) \right] - 1 \right\} \nonumber \\
 +&2\text{Re} \Bigg( \frac{i\lambda_{\text{c}}\left[x_s + i \lambda_{\text{c}} (1-2\frac{\bar v}{v}) \right]\left[x_s+i\lambda_{\text{c}}(1-\frac{\bar v}{v})\right]}{ 2 x_s\left[x_s + i\lambda_{\text{c}}(1 +  \frac{\bar v}{v})\right] \left[x_s+i\lambda_{\text{c}}\right]} \left\{\exp\left[ - \left(\omega_{\text{f}} + \omega_{\text{i}} \right)\left(\frac{\lambda_{\text{c}}}{\bar v} - i\frac{x_s}{\bar v} \right) \right] -1 \right\} \Bigg)       \Bigg].
\end{align}
Only the contribution given by Eq.~(\ref{eq:p_excited_h}) remains in the limit $\omega_{\text{i}}\gg \bar v / \lambda_{\text{c}}$, in which injected and excited electrons can be distinguished by their energy~\footnote{An asymmetry between the distribution of excited electrons in Eq.~(\ref{eq:p_excited_e}) and holes in Eq.~(\ref{eq:p_excited_h}) causes a violation of particle number conservation for excited charge carriers. This violation may stem from the approximations employed to derive the Green's functions given by Eq.~(9), and vanishes in the scaling limit, $x/\lambda \to \infty$ and $\nu/2\pi v \to 0$, at $x_\text{s}/\lambda = x\nu/ \lambda 2\pi v = \text{const}$.}.

\section{Derivation of the two-velocity model}

Here we state the derivation of the Green's functions in the two-velocity model, Eq.~(9) of the main text. The starting point is the exact expression $S^{\text{(exp)}\pm}$ in Eq.~(8) of the main text, obtained for the exponentially decaying interaction stated in Eq.~(7). We have
\begin{align}
\label{eq:S_two_v}
 S^{\text{(exp)}\pm}(x,t) = \sum_{k=1}^{\infty} \frac{1}{k} \left[ \frac{\frac{\nu}{2\pi} t}{x -vt\pm i\lambda k} \right]^k &\approx  \sum_{k=1}^{\infty} \frac{1}{k} \left[ \frac{\frac{\nu}{2\pi} t}{x -vt\pm i\lambda_{\text{c}} } \right]^k \nonumber \\
&= \sum_{k=1}^{\infty} \frac{1}{k} \left[ -\left( \frac{x -\left(v+ \frac{\nu}{2\pi}  \right) t \pm i\lambda_{\text{c}}}{x -vt\pm i\lambda_{\text{c}} } -1 \right) \right]^k \nonumber \\
&= \ln\left( \frac{x-vt \pm i\lambda_{\text{c}}}{x-\bar v t \pm i\lambda_{\text{c}}} \right),
\end{align}
where in the first line we introduced the approximation $\lambda k \to \lambda_{\text{c}}$, with the same effective screening length $\lambda_{\text{c}}$ for all $k$. In the last equality, we insert the definition of the plasmon velocity at the Fermi level, $\bar v  = v + \frac{\nu}{2\pi}$. Insertion of Eq.~(\ref{eq:S_two_v}) into Eq.~(4) in the main text produces Eq.~(9), and completes the derivation.

\section{Full resummation of high injection energy fermion/plasmon perturbation theory for arbitrary bosonic dispersion}

In this section, we show how the electron distribution $p_{\text{sc}}$ is obtained in the limit of high injection energies $\omega_{\text{i}}$. In this limit, the high energy electron can be treated separately from plasmons in the Fermi sea, such that the spatial representation of the interaction part of the Hamiltonian in Eq.~(1) in the main text can be written as
%
%**************** interaction Hamiltonian  **************************
\begin{equation}
\label{eq:Hint}
H_{\rm int} \ = \ \int dx \, dx^\prime \hat{\rho}_{\rm el}(x)  \nu(x - x^\prime) \hat{\rho}_{\rm bos}(x^\prime)\ .
\end{equation}
%*********************************************************************
%
This Hamiltonian describes interactions between the densities $\hat{\rho}_{\rm el}(x)$ of the injected electron and $\hat{\rho}_{\rm bos}(x^\prime)$ of plasmons in the Fermi sea.
Due to the fact that the velocity of high-energy electrons is given by $v$ independent of their momentum, the classical electron trajectory is independent of the energy of the electron. Setting injection position and injection time to zero, the semiclassical electron density is 
given by 
%
%*************** semiclassical electron density  ********************
\begin{equation}
\label{eq:rhoel}
\hat{\rho}_{\rm el} \ = \ \delta(x - v t) \ , \ \ \  0 \leq t \leq t_f \ .
\end{equation}
%****************************************************************
%
Here, $t_f$ is chosen such that $x_f = v t_f$, and $x_f$ denotes the position of the detector quantum dot at which the electron is extracted from the channel. Based on Eqs.~(\ref{eq:Hint}) and~(\ref{eq:rhoel}), the time evolution operator in the interaction picture becomes (setting $\hbar = 1$)
%
%*******************  semiclassical time evolution  **********************
\begin{equation}
\label{eq:U}
\hat{U}
(t)  \ = \ \hat{T} e^{- i \int_0^t dt^\prime \int dx^\prime \nu(v t^\prime - x^\prime) \hat{\rho}_{\rm bos}(x^\prime,t^\prime)}.
\end{equation}
%**********************************************************************
% 

We now set out to obtain the electron distribution $p_{\text{sc}} = \sum_{N=0}^{\infty} p_{\text{sc}}^{(N)}$  as a sum of the distribution $p_{\text{sc}}^{(N)}$ in which electrons, injected at $\omega_{\text{i}}$, excite $N$ plasmons in the Fermi sea, before being detected energy at $\omega_{\text{f}}$. From Fermi's Golden rule, we have
\begin{align}
\label{eq:pN}
  p_{\text{sc}}^{(N)}(x,\omega_{if})  & =  2 \pi \sum_{q_1, q_2, \dots,q_N}    \delta\left(\omega_{q_1} + \omega_{q_2} + ...+ \omega_{q_N} - \omega_{if} \right) \left| T^{(N)}\left(x; \{q_i\}\right) \right|^2,
\end{align}  
in which $\omega_{\text{if}}$ denotes the difference $\omega_{\text{i}}-\omega_{\text{f}}$  between injection and detection energies, $\omega_q = vq(1 + \nu_q/2 \pi v)$ the plasmon dispersion, and	
%
%************************  N plasmon amplitude   ********************
\begin{equation}
\label{eq:TN1}
T^{(N)}(x; \{q_i\}) \ = \ \langle \{ q_i\}| \hat{U}(x/v) |0\rangle  
\end{equation}
%**********************************************************************
%
denotes the transition amplitude from the ground state $|0\rangle$ of the plasmon system, to the state 
%
%**********************  plasmon number eigenstates ***********
\begin{equation}
|\{ q_i\}\rangle \ = \ \left(\prod_{i=1}^N b_{q_i}^\dagger \right) |0\rangle 
\end{equation}
%**************************************************************
%
which features $N$ excited plasmons with momenta $\{ q_i\}$. 
Here, $b_q^\dagger$ is the plasmon creation operator, which obeys the commutation relation $[b_q , b^\dagger_{q^\prime}] = \delta_{q,q^\prime}$. 

In the limit of high injection energies, the greater electron Green's function can be expressed as \cite{Neuenhahn2009supp}
$G_{\text{sc}}^>(x,\omega_{\text{i}}) \ = \ G_0^>(x,\omega_{\text{i}}) \ \langle 0 | \hat{U}(x/v) |0\rangle$, and we identify [see also right below Eq.~(\ref{eq:pHighE})]
%
%********************  greater electron Green function   *****************
\begin{equation}
\label{greater_greenfunction.eq}
\langle 0 | \hat{U}(x/v) |0\rangle =\exp \left[S^>\left(x,\frac{x}{v} \right) \right].
\end{equation}
%*********************************************************************

\subsection{Semiclassical $N$-plasmon amplitudes}

In order to evaluate the semiclassical amplitudes $T^{(N)}(x; \{q_i\})$ defined in Eq.~(\ref{eq:TN1}) exactly, we need to evaluate expectation values of the type 
%
%******************  example expectation values  **************
\begin{equation}
\label{eq:expectation}
\langle 0| b_{q_N} \cdot .... \cdot  b_{q_1} e^{-i\sum_q(\alpha_q b_q + \alpha_q^* b_q^\dagger)}|0\rangle \ .
\end{equation}
%*****************************************************************
By expanding the exponential in Eq.~(\ref{eq:expectation}) into a power series, this can be reduced to expectation values of the type 
%
%******************* power series expectation values   **************
\begin{equation}
\label{eq:INn}
I(N,n) \ = \ \langle 0| b_{q_N} \cdot .... \cdot  b_{q_1} {(-i)^n \over n!} \left[\sum_q(\alpha_q b_q + \alpha_q^* b_q^\dagger)\right]^n |0\rangle \ .
\end{equation}
%************************************************
%
Using Wick's theorem to perform partial contractions, these expectation values can be rewritten as
%
%***********************  power series Wick's theorem *****************************
\begin{align}
\label{eq:INn2}
I(N,n) 
&=    (-i)^N  \langle 0| b_{q_N} \cdot .... \cdot  b_{q_1} \left[\sum_q(\alpha_q b_q + \alpha_q^* b_q^\dagger)\right]^N |0\rangle 
\left(\begin{array}{c}  n \\ N \end{array} \right)
{(-i)^{n-N}\over n!} \langle0|  \left[\sum_q(\alpha_q b_q + \alpha_q^* b_q^\dagger)\right]^{n-N} |0\rangle \nonumber \\
&=  {(-i)^N \over N!} \langle 0| b_{q_N} \cdot .... \cdot  b_{q_1} \left[\sum_q(\alpha_q b_q + \alpha_q^* b_q^\dagger)\right]^N |0\rangle 
{(-i)^{n-N}\over (n-N)!} \langle0|  \left[\sum_q(\alpha_q b_q + \alpha_q^* b_q^\dagger)\right]^{n-N} |0\rangle \nonumber \\
&=  T^{(N)}_{\rm pert} \ {(-i)^{n-N}\over (n-N)!} 
\langle0|  \left[\sum_q(\alpha_q b_q + \alpha_q^* b_q^\dagger)\right]^{n-N} |0\rangle \ ,
\end{align}
where we defined
\begin{align}
\label{eq:TNpert}
T^{(N)}_{\rm pert} := {(-i)^N \over N!} \langle 0| b_{q_N} \cdot .... \cdot  b_{q_1} \left[\sum_q(\alpha_q b_q + \alpha_q^* b_q^\dagger)\right]^N |0\rangle \ .
\end{align}
%************************************************************************************
%
Using Eq.~(\ref{eq:INn2}), we can resum the exponential in Eq.~(\ref{eq:expectation}) to obtain
%
%*********************** semiclassical N plasmon amplitude evaluated  ************
\begin{eqnarray} 
\label{eq:TN}
T^{(N)}(x,\{q_i\}) & = & T^{(N)}_{\rm pert} (x,\{q_i\}) \cdot \langle 0 | \hat{U}(x/v) |0\rangle \nonumber \\
& = & T^{(N)}_{\rm pert} (x,\{q_i\}) \cdot \exp \left[S^>\left(x,\frac{x}{v} \right) \right],
\end{eqnarray}
%********************************************************************************
%
where we used Eq.~(\ref{greater_greenfunction.eq}).
Due to the fact that the perturbative $N$-plasmon transition amplitudes Eq.~(\ref{eq:TNpert}) factorize in the form 
%
%*********************** factorization of perturbative plasmon matrix elements   ****************
\begin{equation}
\label{eq:TN2}
T^{(N)}_{\rm pert}(x;\{q_i\}) \ = \ \frac{1}{N!} \prod_{i=1}^N T^{(1)}_{\rm pert}(x;q_i),
\end{equation}
%*************************************************************************************************
%
a closed form evaluation of $T^{(N)}(x;\{q_i\})$ is possible.

\subsection{Perturbation theory}

To make connection with the perturbative calculation of T-matrix elements considered so far, we need to expand the time evolution operator Eq.~(\ref{eq:U}) in powers of the interaction Hamiltonian in Eq.~(\ref{eq:Hint}), and evaluate Eqs.~(\ref{eq:TN}) and~(\ref{eq:TN2}). For the perturbative one plasmon amplitude ($N=1$ in Eq.~(\ref{eq:TNpert})), we obtain
%
%********************** perturbative one plasmon amplitude  ***************
\begin{equation}
T^{(1)}_{\rm pert}(x;q) \ = \ \langle 0| \hat{b}_q (-i) \int_0^{x/v} dt^\prime \int dx^\prime \nu(v t^\prime - x^\prime) 
\hat{\rho}_{\rm bos}(x^\prime,t^\prime)  | 0 \rangle \ \ .
\label{oneplasmon_pert.eq}
\end{equation}
%*********************************************************************************
%
We now express the plasmon distribution in terms of the displacement field as $\hat{\rho}(x,t) = (1/2 \pi) \partial_x \hat{\varphi}(x,t)$, with
%
%**************************  plasmon field operator ******************
\begin{equation}
\label{eq:expansion}
\hat{\varphi}(x,t) \ = \ \sum_{q >0} \sqrt{2 \pi \over q L} \left(\hat{b}_q e^{i q x - i \omega_q t}  + \hat{b}_q^\dagger e^{- i q x + i \omega_q t} \right) \ ,
\end{equation}
%*******************************************************************
%
where $\omega_q = q(v + \nu_q/2 \pi)$ denotes the plasmon dispersion relation, and $\nu_q = \int_{-\infty}^{\infty} dx\, \nu(x) e^{- i q x}$ denotes the interaction matrix element in $q$-space. Upon insertion of the expansion in Eq.~(\ref{eq:expansion}) into Eq.~(\ref{oneplasmon_pert.eq}), we obtain
%
%********************** evaluation of perturbative on plasmon amplitude   **************
\begin{eqnarray}
\label{oneplasmon_pert_eval.eq}
T^{(1)}_{\rm pert}(x;q) & = &  - i \sqrt{ 2 \pi \over q L} \left(1 - e^{i q x {\nu_q \over 2 \pi v}}\right).
\end{eqnarray}
%********************************************************************************************
%

\subsection{Resummation of semiclassical N-plasmon amplitudes}

The results of the previous sections allow us to resum the full series that determines the electron distribution in the limit of high injection energies. Insertion of Eq.~(\ref{eq:TN}) into Eq.~(\ref{eq:pN}) leads to
%
%************************ N plasmon rate  *************************
\begin{align}
\label{eq:pN2}
  p_{\text{sc}}^{(N)}(x,\omega_{\text{if}}) 
  & =  2 \pi \left| \exp \left[S^>\left(x,\frac{x}{v} \right) \right] \right|^2 { 1 \over N!} \sum_{q_1, q_2, \dots,q_N}    \delta\left(\omega_{q_1} + \omega_{q_2} + ...+ \omega_{q_N} - \omega_{if} \right) \left| \prod_{i=1}^N T^{(1)}_{\rm pert}(x;q_i) \right|^2, \nonumber \\
& =  2 \pi \left| \exp \left[S^>\left(x,\frac{x}{v} \right) \right] \right|^2  \int_{-\infty}^\infty {d t \over 2 \pi} e^{i  \omega_{if} t} \ { 1 \over N!} 
\left[ \sum_{q} \left|T^{(1)}_{\rm pert}(x;q) \right|^2 e^{-i\omega_q t}\right]^N.
\end{align}
%******************************************************************
%
Upon insertion of Eq.~(\ref{oneplasmon_pert_eval.eq}), as well as Eq.~(5) from the main text into~(\ref{eq:pN2}),
the derivation is completed after summation over all $N$, which produces Eq.~(10) in the main text.
Written in terms of Green's functions, Eq.~(10) reads
\begin{align}    
\label{eq:pHighE}
  &p_{\text{sc}}(x,\omega_{\text{i}},\omega_{\text{f}}) \stackrel{\omega_{\text{i}} \gg \frac{\bar v}{\lambda},\, \omega_{\text{if}}}=  \frac{ v^2}{2\pi} G_{\text{sc}}^{-}\left(x, -\omega_\text{i} \right)G_{\text{sc}}^{+}\left(x,\omega_\text{i} \right) \nonumber \\
  & \qquad \qquad \times  \int_{-\infty}^{+\infty} \! dt_0  e^{i \omega_{\text{if}} t_0 } \frac{G^-\left(0, -t_0 \right)G^+\left(0, t_0 \right) }{G^-\left(x, -t_0 \right)G^+\left(x,t_0 \right) },
\end{align}
cf.~\cite{Takei2010supp} and Eq.~(6) in the main text. Here $G^{\pm}_{\text{sc}}(x,\omega) = G_{0}^{\pm}\left(x, \omega\right) \exp \left[ S^{\pm}(x,x/v)\right]$ denotes the GF's high energy limit discussed by~[\onlinecite{Neuenhahn2008supp}], which contains information about elastic transfer.

\section{Cancellation of $N>1$ plasmon distributions with higher order $x_{\text{s}}$ terms of the $N=1$ plasmon distribution, in the two-velocity model}

\begin{figure}[t]
    \includegraphics[width=.65\textwidth]{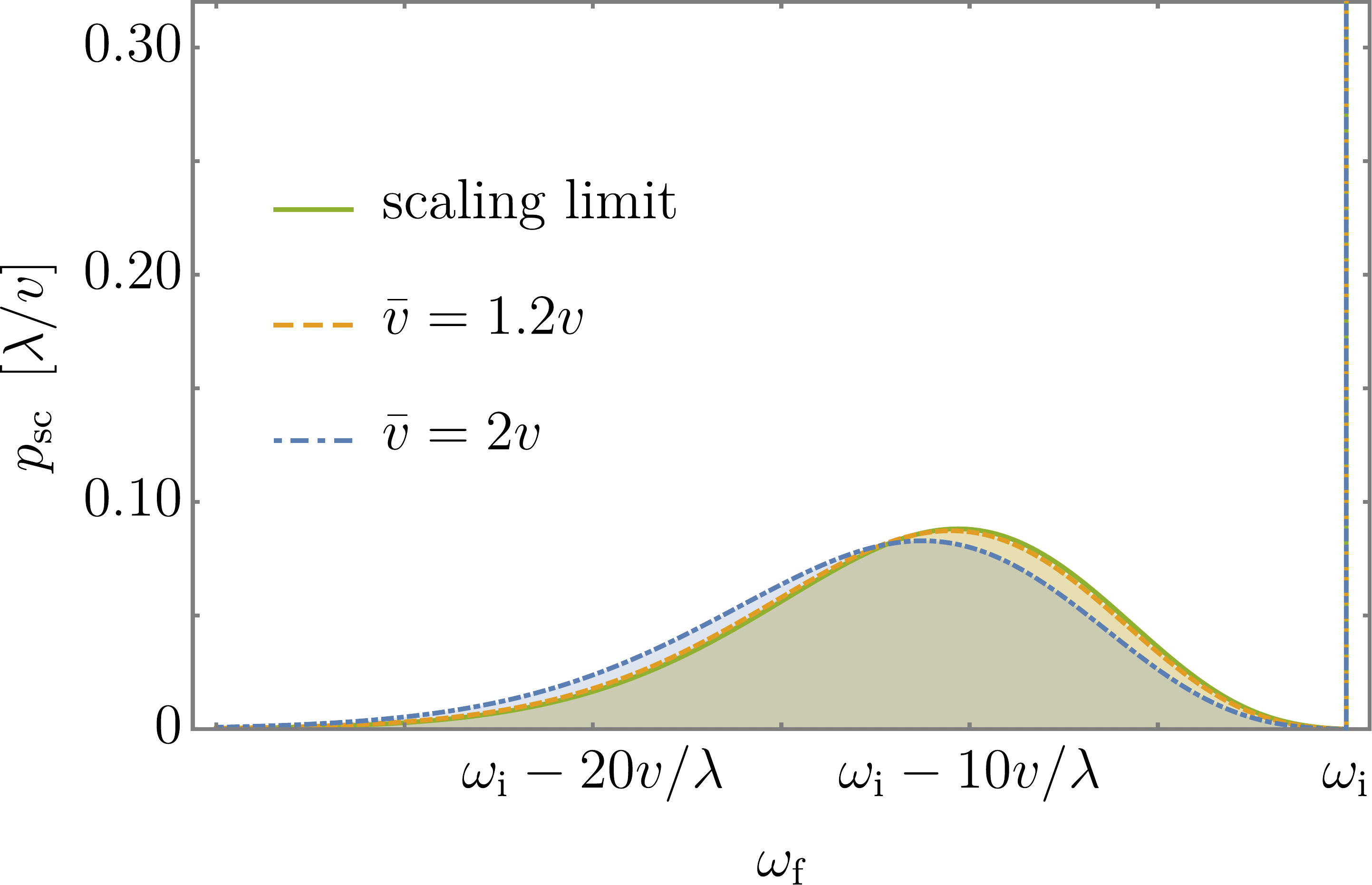}
    \caption{\label{fig:scaling_limit} Electron distribution $p_{\text{sc}}$ at $x_{\text{s}} = 11 \lambda$, evaluated for interactions that decay exponentially in momentum space, in the scaling limit (green) as well as for weak (yellow, $\bar v = 1.2v$) and stronger (blue, $\bar v = 2v$) interactions. The scaling limit result, which coincides with the green curve in Fig.~3 in the main text, shows excellent agreement with the result for weak interactions. Stronger interactions stretch the distribution towards lower detection energies.
    }
\end{figure}

For the Green's functions given by Eq.~(9) which feature the velocity of the injected electron $v$ and one velocity $\bar v$ for all plasmons excited from the Fermi sea, the distribution $p_{\text{sc}}^{(1)}$ that corresponds to the excitation of one plasmon in the Fermi sea  is given by
\begin{align}
\label{eq:psc1}
p_{\text{sc}}^{(1)} &= \frac{\lambda_{\text{c}}^2}{x_{\text{s}}^2 + \lambda_{\text{c}}^2} \frac{4}{\omega_{\text{if}}} \sin^2 \left(\frac{\omega_{\text{if}} x_{\text{s}}}{2\bar v} \right)  \exp \left( -\omega_{\text{if}} \frac{\lambda_{\text{c}}}{\bar v}   \right)\nonumber \\
&=  \frac{x_{\text{s}}^2}{x_{\text{s}}^2 + \lambda_{\text{c}}^2} \frac{\lambda_{\text{c}}^2}{\bar v^2} \omega_{\text{if}} \left[1 + \mathcal{O}\left( \frac{x_{\text{s}}^2}{\lambda_{\text{c}}^2} \right) \right]  \exp \left( -\omega_{\text{if}} \frac{\lambda_{\text{c}}}{\bar v}   \right).
\end{align}
Comparison of Eq.~(\ref{eq:psc1}) with the fully resummed distribution in Eq.~(2), obtained for the the two-velocity model in Eq.~(9), shows that the $\mathcal{O} \left(1\right)$ contribution which comes with the second line of Eq.~(\ref{eq:psc1}) already coincides with the fully resummed expression.  This means that the higher order terms $\mathcal{O}\left( x_{\text{s}}^2/\lambda_{\text{c}}^2 \right)$ 
 in the square brackets in Eq.~(\ref{eq:psc1}) cancel all distribution terms that feature more than one plasmon, $p_{\text{sc}}^{(N)}$, $N>1$, within the framework of this model. For general screened interaction matrix elements $\nu_q$ in Fourier space, which give rise to a continuum of plasmon velocities via $\omega_q$, this cancellation is no longer perfect.

\section{Scaling limit}
\label{sec:scaling_limit}

We discuss the scaling limit as defined in the main text, in which $x/\lambda \to \infty$ and $\nu/2\pi v \to 0$, at $x_\text{s}/\lambda = x\nu/ \lambda 2\pi v = \text{const}$ (that is, upon neglecting all interaction terms $\nu/2\pi v$ that do not appear in a product with $x/\lambda$).
The scaling limit neglects effects of interactions other than the finite spatial spread of the wave packet in the vicinity of the detection point.
In this limit, the distribution given by Eq.~(10) in dimensionless form, $v p_{\text{sc}} /\lambda$, depends only on the dimensionless distance $x_{\text{s}}/\lambda$ and energy loss $\omega_{\text{if}}\lambda/v$, and allows, e.g.~for the exponentially decaying interaction in Eq.~(7), for efficient numerical evaluation even at large values of $x_{\text{s}}/\lambda$.
 Figure~\ref{fig:scaling_limit} shows the distribution $p_{\text{sc}}$, Eq.~(10) in the main text, evaluated for such interactions that decay exponentially in momentum space, at $x_{\text{s}} = 11 \lambda$ in the scaling limit (green curve, which coincides with the green curve in Fig.~3 in the main text), as well as for interaction strengths $\nu /2\pi v = 0.2$ and $\nu /2\pi v = 1$, such that $\bar v =1.2v$ and $\bar v =2v$, respectively.
The scaling limit result shows excellent agreement with the result at relatively weak interactions $\bar v =1.2v$. 
 Note that, similarly to the high injection energy electron distribution obtained from the two-velocity model [cf.~Eq.~(2) in the main text], stronger interactions stretch the distribution towards lower detection energies $\omega_{\text{f}}$.